\documentstyle{aipproc}

\begin{document}

\title{Bloch--Wilson Hamiltonian \\
and a Generalization of \\
the Gell-Mann--Low Theorem\footnote{This work was supported in part by
Conacyt grant 3298P--E9608 and the Coordinaci\'on de la Investigaci\'on
Cient\'\i fica of the Universidad Michoacana de San Nicol\'as de Hidalgo.}}

\author{Axel Weber\thanks{e--mail: axel@io.ifm.umich.mx}}

\address{Instituto de F\'\i sica y Matem\'aticas, Universidad Michoacana de
San Nicol\'as de Hidalgo, \\
Edificio C--3 Cd.\ Universitaria, A. P. 2--82, 58040 Morelia, Michoac\'an,
Mexico}

\maketitle

\begin{abstract}
The effective Hamiltonian introduced many years ago by Bloch and generalized
later by Wilson, appears to be the ideal starting point for Hamiltonian
perturbation theory
in quantum field theory. The present contribution derives the Bloch--Wilson
Hamiltonian from a generalization of the Gell-Mann--Low theorem, thereby
enabling a diagrammatic analysis of Hamiltonian perturbation theory in this
approach.
\end{abstract}

The presently available techniques for calculations in quantum field theory
reflect the dominance of scattering processes for the experimental
exploration of the physics of elementary particles. The single most important
technique is beyond doubt Lagrangian perturbation theory, the explicit
covariance of which has historically played an important r\^ole in the
implementation of
the renormalization program. This in turn was the crucial ingredient for
converting the formal expressions of Lagrangian perturbation theory into
predictions for measurable quantities. On the other hand, the identification
of physical states defined as eigenstates of the Hamiltonian and the Hilbert
space they span, becomes a complicated task in this approach, which
is exemplified by the serious problems arising in the solution of the
Bethe--Salpeter equation. In short, Lagrangian perturbation theory is
primarily a theory of processes as opposed to a theory of states.

This contribution is concerned with the development of a theory of states,
establishing efficient techniques for Hamiltonian perturbation theory.
Apart from the possibility of gaining a new perspective on the foundations
of quantum field theory, this approach appears to be natural for the
description of hadronic structure and of bound state phenomena in general.
In a very general setting, consider the problem of solving the Schr\"odinger
equation
\begin{equation}
H | \psi \rangle = E | \psi \rangle
\label{axel:schr}
\end{equation}
for the state $| \psi \rangle$. The Hamiltonian is supposed to be
decomposable into a ``free'' and an ``interacting'' part, $H = H_0 + H_I$,
where the eigenstates of $H_0$ are explicitly known and span the full Hilbert
space (or Fock space)
$\cal F$, which we picture as a direct sum of free $n$--particle subspaces
$(n \ge 0)$. The eigenstates of $H$ are expected to be representable as
(infinite) linear combinations of the eigenstates of $H_0$, hence the
Schr\"odinger equation (\ref{axel:schr})
can be written in a Fock space basis, where in general an infinite
number of $n$--particle subspaces are involved. The problem in this
generality is obviously too difficult to be solved in practice.

Restricting attention momentarily to the vacuum state, the Gell-Mann--Low
theorem \cite{axel:GL} states that the free (Fock space) vacuum evolves
dynamically into the physical vacuum as $H_0$ turns adiabatically into
$H$. Explicit expressions can then be given for the physical vacuum state
and its energy in terms of the free $n$--particle states and their energies
in the form of a perturbative series. It is natural to ask whether it is
possible to generalize the theorem to the case where the perturbative vacuum
is replaced by a linear subspace $\Omega$ of $\cal F$ consisting of
eigenspaces of
$H_0$, i.e.\ $H_0 \Omega \subseteq \Omega$, the simplest non--trivial
example being the free two--particle subspace of $\cal F$. When the
interaction $H_I$ is switched on adiabatically, one may expect that $\Omega$
evolves into the suspace of interacting physical two--particle states,
where now different eigenstates of $H_0$ are allowed to mix during the
adiabatic process. If this expectation comes true, the determination
of the physical two--particle states may be reduced to a problem within the
free two--particle subspace, thus dramatically reducing the number of
degrees of freedom to be considered and converting the problem into a (at
least numerically) solvable one.

Couched into mathematical jargon, what one is looking for is a map $U_{BW}$
from $\Omega$ to a direct sum of eigenspaces of $H$, i.e.\ $H U_{BW} \Omega
\subseteq U_{BW} \Omega$, where $U_{BW}$ is expected to be related to the
adiabatic evolution operator. One would then hope that $U_{BW}$ induces a
similarity transformation, so that the problem of diagonalizing $H$ in $U_{BW}
\Omega$ is equivalent to diagonalizing $H_{BW} := U_{BW}^{-1} H U_{BW}:
\Omega \to \Omega$, which in the example above is equivalent to a relativistic
two--particle Schr\"odinger equation. The simplest (but not unique) choice
for $U_{BW}^{-1}: U_{BW} \Omega \to \Omega$ is the orthogonal projector $P$
to $\Omega$,\footnote{That the choice of $P$ for the
similarity transformation is not unreasonably
simple is suggested by phenomenology: even in the highly non--perturbative
situation of low--energy QCD the physical hadrons can be associated with
a specific content of constituent quarks (and thus with an element of the
free two-- or three--particle subspace).} hence we will look for an operator
$U_{BW}$ in $\Omega$ with
\begin{equation}
P U_{BW} = P = {\bf 1}_\Omega \:.
\label{axel:proj}
\end{equation}
Eq.\ (\ref{axel:proj}) implies in turn the injectivity of $U_{BW}$, hence
also $U_{BW} P = {\bf 1}$ in $U_{BW} \Omega$. Together with $H U_{BW} \Omega
\subseteq U_{BW} \Omega$ one then has that
\begin{equation}
({\bf 1} - U_{BW} P) H U_{BW} = 0 \;\: \mbox{in} \;\: \Omega \:.
\label{axel:wilson}
\end{equation}
Eqs.\ (\ref{axel:proj}) and (\ref{axel:wilson}) together in fact characterize
$U_{BW}$: (\ref{axel:wilson}) implies $H U_{BW} \Omega = U_{BW} (P H U_{BW}
\Omega) \subseteq U_{BW} \Omega$. Consequently, $H |_{U_{BW} \Omega}$ is
diagonalizable, and by (\ref{axel:proj}) it is a similarity transform of
$H_{BW}$.

Remarkably, Eqs.\ (\ref{axel:proj}) and (\ref{axel:wilson}) also determine
$U_{BW}$ uniquely, at least within the perturbative regime. To see this,
rewrite (\ref{axel:wilson}) as
\begin{eqnarray}
H_I U_{BW} - U_{BW} P H_I U_{BW} &=& U_{BW} P H_0 U_{BW} - H_0 U_{BW}
\nonumber \\
&=& U_{BW} H_0 - H_0 U_{BW} \:,
\label{axel:bloch}
\end{eqnarray}
where I have used $P H_0 U_{BW} = H_0 P U_{BW} = H_0$. Now consider
the matrix element of (\ref{axel:bloch}) between $\langle u |$ and
$| k \rangle$, where $| k \rangle \in \Omega$ and $| u \rangle \in
\Omega^\perp$ (the orthogonal complement of $\Omega$ in $\cal F$) are
eigenstates of $H_0$ with eigenvalues $E_k$ and $E_u$, respectively,
\begin{equation}
\langle u | H_I U_{BW} - U_{BW} P H_I U_{BW} | k \rangle = (E_k - E_u)
\langle u | U_{BW} | k \rangle \:.
\end{equation}
It then follows that
\begin{eqnarray}
U_{BW} &=& P + ({\bf 1} - P) U_{BW} P \nonumber \\
&=& P + \int_\Omega dk \int_{\Omega^\perp} du \, | u \rangle \langle u |
U_{BW} | k \rangle \langle k | \nonumber \\
&=& P + \int_\Omega dk \int_{\Omega^\perp} du \, | u \rangle \frac{\langle u |
H_I U_{BW} - U_{BW} P H_I U_{BW} | k \rangle}{E_k - E_u} \langle k | \:,
\label{axel:pert}
\end{eqnarray}
where I have taken $k$ and $u$ to label the eigenstates of $H_0$ in
$\Omega$ and $\Omega^\perp$, respectively. Eq.\ (\ref{axel:pert}) can be
solved iteratively to obtain $U_{BW}$ as a power series in $H_I$. It should
be emphasized, however, that the individual terms in the series are not
guaranteed to give convergent expressions (let alone the series as a whole).
This depends, among other things, on the choice of $\Omega$.

Eqs.\ (\ref{axel:proj}) and (\ref{axel:wilson}) have been used for the
characterization of $U_{BW}$ before, first by Bloch \cite{axel:CB} in the
context of degenerate quantum mechanical perturbation theory, and later by
Wilson \cite{axel:KGW} for the formulation of a non--perturbative
renormalization group in Minkowski space.
In practical applications, one will calculate $U_{BW}$ to a certain order
in the iterative expansion of (\ref{axel:pert}) and solve the Schr\"odinger
equation for the corresponding Hamiltonian $H_{BW} = P H U_{BW}$. Its solution
yields an approximation to the eigenvalues of $H |_{U_{BW} \Omega}$ (the
eigenvalues are invariant under similarity transformations) and also to
the eigenstates via $| \psi \rangle = U_{BW} | \phi \rangle$ where
$| \phi \rangle$ are the eigenstates of $H_{BW}$. The solutions will in
general also include bound states (e.g., if $\Omega$ is the free two--particle
subspace), in contrast to Lagrangian perturbation theory. The reason for this
difference is that
although in the present formalism $H_{BW}$ is determined perturbatively,
the corresponding Schr\"odinger equation can be solved exactly (at least to
arbitrary precision with numerical methods). This is somewhat analogous to
the Bethe--Salpeter equation, but avoids the conceptual problems associated
with the latter. In this context, it is worth mentioning that the
normalizability of the free two--particle component $| \phi \rangle = P
| \psi \rangle$ gives a natural criterium for the ``boundedness'' of the
state $| \psi \rangle$, although the latter may not be normalizable in the
Hilbert space sense.

The formulation presented so far has two important shortcomings: first, the
terms in the perturbative series following from (\ref{axel:pert}) are
not well--defined
in the case of vanishing energy denominators, and a consistent prescription
is at least not obvious from (\ref{axel:wilson}) or (\ref{axel:pert}).
Second, it is not a priori clear how to translate the terms in the
perturbative series into diagrams. A
diagrammatic formulation, however, is expected to be at least helpful, if
not imperative, for the investigation of such important properties as
renormalizability and Lorentz and gauge invariance at finite orders of the
expansion, as well as for practical applications of the formalism.

In search of an alternative characterization of $U_{BW}$, I will now return
to the idea of the adiabatic evolution. Consider the adiabatic evolution
operator from $t = -\infty$ to $t=0$,
\begin{eqnarray}
U_{\epsilon} &=& T \exp - i \int_{-\infty}^0 dt \,
e^{-\epsilon |t|} H_I(t) \nonumber \\
&=& \sum_{n=0}^\infty \frac{(-i)^n}{n!}
\int_{-\infty}^0 dt_1 \cdots \int_{-\infty}^0 dt_n \, e^{-\epsilon (|t_1| +
\ldots + |t_n|)} \, T [ H_I (t_1) \cdots H_I (t_n) ] \:,
\label{axel:evol}
\end{eqnarray}
where
\begin{equation}
H_I(t) = e^{i H_0 t} H_I \, e^{-i H_0 t}
\end{equation}
is the usual expression in the interaction picture and $T$ stands for the
decreasing time ordering operator. Then the following theorem holds:
\smallskip

\noindent
{\bf Generalized Gell-Mann--Low Theorem.}
{\em With the notations introduced before, suppose that the operator $U_{BW}
:= \lim_{\epsilon \to 0} \, U_\epsilon (P U_\epsilon P)^{-1}$ exists in
$\Omega$. Then it has the properties $P U_{BW} = P$ and
$({\bf 1} - U_{BW} P) H U_{BW} = 0$ in $\Omega$.}
\smallskip

{\it Remarks.}\hspace{1ex}
We thus have an explicit expression for $U_{BW}$ in terms of the adiabatic
evolution operator. Given that $P U_\epsilon P$ is
always formally invertible as a power series in $H_I$, the implications of the
theorem rest on the existence of the limit $\epsilon \to 0$ of
$U_\epsilon (P U_\epsilon P)^{-1}$, which in turn depends on the choice of
$\Omega$.

{\it Proof.}\hspace{1ex}
The property $P U_{BW} = P$ follows directly from the definition of $U_{BW}$.
The first part of the proof of $({\bf 1} - U_{BW} P) H U_{BW} = 0$ is
identical to the original Gell-Mann--Low proof \cite{axel:GL} and will not be
reproduced here. It establishes by manipulation of the series (\ref{axel:evol})
for $U_\epsilon$ that (before taking the limit $\epsilon \to 0$)
\begin{equation}
H U_\epsilon = U_\epsilon H_0 + i \epsilon g \frac{\partial}{\partial g}
U_\epsilon \:,
\label{axel:adiab}
\end{equation}
where $H_I$ is assumed to be proportional to some ``coupling constant'' $g$.

Now choose any $| \phi \rangle \in \Omega$. Eq.\ (\ref{axel:adiab}) implies
\begin{equation}
H U_\epsilon (P U_\epsilon P)^{-1} | \phi \rangle = U_\epsilon H_0
(P U_\epsilon P)^{-1} | \phi \rangle + i \epsilon \left( g
\frac{\partial}{\partial g} U_\epsilon \right) (P U_\epsilon P)^{-1}
| \phi \rangle \:.
\label{axel:ident}
\end{equation}
It follows that
\begin{eqnarray}
\lefteqn{H U_\epsilon (P U_\epsilon P)^{-1} | \phi \rangle - i \epsilon
g \frac{\partial}{\partial g} \left( U_\epsilon (P U_\epsilon P)^{-1} \right)
| \phi \rangle} \hspace{1cm} \nonumber \\
&=& U_\epsilon H_0 (P U_\epsilon P)^{-1} | \phi \rangle + i \epsilon
U_\epsilon (P U_\epsilon P)^{-1} \left( P g \frac{\partial}{\partial g}
U_\epsilon \right) (P U_\epsilon P)^{-1} | \phi \rangle
\label{axel:first} \\
&=& U_\epsilon (P U_\epsilon P)^{-1} P H U_\epsilon (P U_\epsilon P)^{-1}
| \phi \rangle \:,
\label{axel:second}
\end{eqnarray}
where in going from (\ref{axel:first}) to (\ref{axel:second}) Eq.\
(\ref{axel:ident}) has been used again, multiplied by $U_\epsilon (P
U_\epsilon P)^{-1} P$ from the left, and $P$ has been inserted to the left of
$H_0$, which is possible due to $H_0 \Omega \subseteq \Omega$. Taking the
limit $\epsilon \to 0$, we have $H U_{BW} | \phi \rangle = U_{BW} P H U_{BW}
| \phi \rangle$,
which proves the theorem. In taking the limit, the existence of the
$g$--derivative of $U_{BW}$ in $\Omega$ has been assumed. Incidentally, this
assumption implies that the expression $U_\epsilon (g \, \partial/ \partial g)
(P U_\epsilon P)^{-1} | \phi \rangle$ is in general {\it divergent} in the
limit $\epsilon \to 0$, since $H U_\epsilon (P U_\epsilon P)^{-1}
| \phi \rangle$ cannot be expected to be equal to  $U_\epsilon H_0
(P U_\epsilon P)^{-1} | \phi \rangle$ in this limit \cite{axel:GL}.
\hfill \rule{3mm}{3mm}

The theorem corroborates the expectation detailed at the
beginning of this contribution. More importantly, the adiabatic formulation
also has the benefit of fixing an $i \epsilon$--prescription for the
energy denominators appearing in the series generated by (\ref{axel:pert}).
Performing the time integrations in $U_\epsilon (P U_\epsilon P)^{-1}$
yields explicitly to second order in $H_I$
\begin{eqnarray}
U_{BW} &=& \int_\Omega dk \, | k \rangle \langle k | +
\int_\Omega dk \int_{\Omega^\perp} du \, | u \rangle \frac{\langle u |
H_I | k \rangle}{E_k - E_u + i \epsilon} \langle k | \nonumber \\
&& {}- \int_\Omega dk \, dk' \int_{\Omega^\perp} du \, | u \rangle
\frac{\langle u | H_I | k' \rangle \, \langle k' | H_I | k \rangle}{(E_k - E_u
+ 2 i \epsilon) (E_{k'} - E_u + i \epsilon)} \langle k | \nonumber \\
&& {}+ \int_\Omega dk \int_{\Omega^\perp} du \, du' \, | u \rangle
\frac{\langle u | H_I | u' \rangle \, \langle u' | H_I | k \rangle}{(E_k - E_u
+ 2 i \epsilon) (E_k - E_{u'} + i \epsilon)} \langle k | + \ldots \:,
\label{axel:epsilon}
\end{eqnarray}
where the limit $\epsilon \to 0$ is understood. The same expression without
the $i \epsilon$--prescription follows from iterating (\ref{axel:pert}).

The second important advantage of the formulation in terms of $U_\epsilon$
is the ready translation into diagrams. The diagrams associated with the
perturbative expansion of $H_{BW}$ turn out to be similar to Goldstone or
time--ordered diagrams, but unlike the latter they do {\it not} combine into
a set of Feynman diagrams. This is essentially due to the fact that the
matrix elements of the effective Hamiltonian $\langle k | H U_{BW} | k'
\rangle$ in general do not vanish if the energies $E_k$ and $E_{k'}$ are
different.

\end{document}